\documentclass[12pt]{article}
\usepackage{amsfonts,amssymb,epsfig,amsmath,times}
\usepackage{float}
\restylefloat{table}
\renewcommand{\text}[1]{#1}

\newcommand{\be}{\begin{equation}}
\newcommand{\ee}{\end{equation}}
\newcommand{\ben}{\begin{displaymath}}
\newcommand{\een}{\end{displaymath}}
\newcommand{\bea}{\begin{eqnarray}}
\newcommand{\eea}{\end{eqnarray}}
\newcommand{\bean}{\begin{eqnarray*}}
\newcommand{\eean}{\end{eqnarray*}}
\newcommand{\nn}{\nonumber \\}
\newcommand{\ba}{\begin{array}}
\newcommand{\ea}{\end{array}}
\newcommand{\bi}{\begin{itemize}}
\newcommand{\ei}{\end{itemize}}

\def\g{\gamma}

\def\G{\Gamma}
\def\d{\delta}

\def\e{\epsilon}

\def\otaula{\begin{tabular}}
\def\ctaula{\end{tabular}}

\def\bnum{\begin{enumerate}}
\def\enum{\end{enumerate}}

\def\CR{\mathcal{R}}
\def\CM{\mathcal{M}}

\def\8M{$\CM_8$}

\def\be{\begin{equation}}
\def\ee{\end{equation}}
\def\G{\Gamma}
\def\g{\gamma}
\def\ei{e^{\underline{i}}}

\def\e1{e^{\underline{1}}}
\def\1u{\underline{1}}
\def\2u{\underline{2}}

\def\0u{\underline{0}}
\def\e{\epsilon}
\def\target{$\CR^{1,1}\times \mathcal{M}_8$ }
\def\target2{$\CR^{1,1}\times \mathcal{M}_8$,}
\def\9G{\G_{\underline{9}}}

\def\1f{f_1^{1/2}}
\def\2f{f_2^{1/2}}
\def\4f{f_4^{1/2}}

\begin{document}
\begin{titlepage}

\vfill

\begin{flushright}
KIAS-P09014
\end{flushright}

\vfill

\begin{center}
   \baselineskip=16pt
   {\Large\bf NR $CFT_3$ duals in M-theory}
   \vskip 2cm
      Eoin \'{O} Colg\'{a}in and Hossein Yavartanoo
         \vskip .6cm
      \begin{small}
      $^1$\textit{Korea Institute for Advanced Study, \\
        Seoul 130-722, Korea}
        \end{small}\\*[.6cm]
\end{center}

\vfill
\begin{center}
\textbf{Abstract}\end{center}

\begin{quote}
We extend the search for supergravity solution duals of non-relativistic $d=3$ CFTs with dynamical exponent $z=2$ to $d=11$ supergravity. We consider the internal space to be an $S^2$ bundle over a product base: $S^2 \times S^2$ and $S^2 \times T^2$. For purely M-theoretic $S^2 \times S^2$, we find only magnetic fluxes preserving two supersymmetries. $S^2 \times T^2$ is far richer admitting, in addition to magnetic fluxes, various non-trivial electric fluxes which break all supersymmetry.
\end{quote}

\vfill

\end{titlepage}
\section{Introduction}
Over recent years the AdS/CFT correspondence has served as a useful holographic tool to explore features of dual relativistic field theories at strong coupling. From its conception as a novel observation, it has gradually gained acceptance as a computational tool and has allowed some important qualitative insights into nuclear physics at RHIC experiments. Natural evolution then leads to application of the AdS/CFT to model conformal field theories which appear in table-top laboratory experiments. Although the microscopic description of these condensed matter systems is not relativistic, there are special quantum critical points which exhibit the full relativistic conformal symmetry. In these special cases, the AdS/CMT allows one to study strongly correlated electrons, superconductors, and the quantum hall effect. 

One may then ask whether holography has a r\^{o}le to play in the understanding of non-relativistic theories away from the special critical points. Such systems are invariant under Galilean transformations $\{H,P\}$, spatial rotations $M$, Galilean boosts $K$ and they also possess a manifest scale invariance $D$. If one considers $x^{+}$ to be the time coordinate and $\mathbf{x}$ to denote the $d$ spatial coordinates, the scaling symmetry acts as
\be
x^{+} \rightarrow \lambda^{z} x^{+}, \quad \mathbf{x} \rightarrow \lambda \mathbf{x},
\ee
where $z$ denotes the dynamical exponent. The algebra of these symmetries, referred to as the Schr\"odinger algebra in $d$ spatial dimensions Sch($d$), is embeddable in the relativistic conformal algebra in $d+2$ spacetime dimensions $O(d+2,2)$, implying that the holographic dual is a $d+3$ dimensional spacetime. Here $z=1$ corresponds to a relativistic theory. For applications of theories with different values of $z$ which describe the onset of (anti)ferromagnetism, etc. in strongly coupled fermion systems in the lab, see \cite{unitarityfermions}. See \cite{Hartnoll:2009sz} for a comprehensive review for the string community. For a flavour of the recent activities being pursued on both sides of the duality, see \cite{Goldberger:2008vg}.

Over the past year, a steady trickle of papers on NR CFT holography have appeared. This paper adds a small contribution to the understanding of the string/M-theory gravity duals of these NR CFTs, the study of which was precipitated by \cite{Son:2008ye,Balasubramanian:2008dm}, when the Schr\"{o}dinger symmetries were realised geometrically. These $d=2$ $z=2$ solutions were subsequently embedded into type IIB string theory \cite{Herzog:2008wg,Maldacena:2008wh,Adams:2008wt}, and featured  spactimes with internal Sasaki-Einstein manifolds $SE_5$. A generalisation of these IIB $SE_5$ solutions appeared in \cite{Hartnoll:2008rs} where the introduction of a harmonic function on $SE_5$ into the external part of the metric allows an interpolating solution that tunes between a solution with two supersymmetries and the IIB solutions above. In \cite{Maldacena:2008wh} and \cite{Gauntlett:2009zw} it was shown how solutions can be obtained as consistent truncations on $SE_5$ from type IIB and $SE_7$ from $d=11$ supergravity to lower-dimensional theories with massive vector fields. In an extension of the latter work \cite{Donos:2009en} it was shown that although the $z=2$ solutions \cite{Herzog:2008wg,Maldacena:2008wh,Adams:2008wt} were non-supersymmetric, among the type IIB solutions \cite{Maldacena:2008wh} with $z \ge 4$ and the $d=11$ supergravity solutions \cite{Gauntlett:2009zw} with $z \ge 3$, there are solutions preserving two supersymmetries.

In this note, as a starting point, we make use of the explicit family of supersymmetric warped $AdS_5$ solutions in $d=11$ supergravity appearing in \cite{Gauntlett:2004zh}. When $M_6$ is a complex manifold, explicit solutions were found which are topologically a two-sphere fibred over a four-dimensional base $\mathcal{B}_4$ which is either K\"ahler-Einstein $KE_4$ or a product of constant curvature Riemann surfaces. Among these products, the $S^2 \times T^2$ product reduces to a IIB solution with internal $SE_5$. Here we will consider an external part of the $d=11$ metric loyal to the isometries of Sch$(2)$, while maintaining the original symmetries of the explicit $M_6$ solutions. 

The structure of this paper is as follows. In section 2 we review the symmetries of the Schr\"{o}dinger group with general dynamical exponent $z$ and explain how they manifest themselves in the metric and the fluxes. In section 3 we analyse the Sch(2) $z=2$ dual with base space $B= S^2 \times S^2$ to determine the nature of the most general solution. This case is compact and when the size of the two spheres are equal, it is $KE_4$, so it exhibits overlap with some of the other explicit solutions. In section 4, we proceed to the uplifted $Y^{p,q}$ solutions with $B=S^2 \times T^2$. Much is already known about these solutions. In section 5 we conclude. 

\section{Preliminaries}
The class of warped $AdS_5 \times_w M_6$ solutions we consider in this paper were initially analysed in \cite{Gauntlett:2004zh} and some of the properties of the $\mathcal{N}=1$ SCFT duals were examined in \cite{Gauntlett:2006ai}. We briefly recap the essential points.  The $d=11$ metric is
\be
ds^{2} = e^{2 \lambda} [ ds^2(AdS_5) + ds^2(M_6)],
\ee
where $\lambda$ is a function of the coordinates on $M_6$ and the four-form flux $G$ is purely magnetic with components only on the $M_6$ space. 

For the special case where $M_6$ admits an integrable almost complex structure and is hence complex, explicit solutions were constructed in \cite{Gauntlett:2004zh}. For the explicit solutions 
\be
ds^{2}(M_6) = ds^{2}(\mathcal{B}_4) + e^{-6 \lambda} \sec^{2} \zeta + \frac{\cos^2 \zeta}{9}(d \psi + P),
\ee
where the Killing vector $\partial_{\psi}$ is related to the R-symmetry of the dual SCFT and $P$ denotes the canonical one-form connection on the base $\mathcal{B}_4$. Topologically these solutions are $S^2$ bundles over a four-dimensional base which is either $KE_4$ or a product of constant curvature Riemann surfaces $\mathcal{C}_1 \times \mathcal{C}_2$, where the curvature of $\mathcal{C}_i$, $k_i \in \{-1,0,1\}$ determines whether $\mathcal{C}_i$ is a hyperbolic space $H^2$, a flat torus $T^2$ or a sphere $S^2$ respectively. 

As the scope of this note is the candidate geometric duals to non-relativistic CFTs in $d=3$, we will consider the above class of solutions but replace $AdS_5$ with the following metric 
\bea
\label{metric1}
ds^{2} = r^2 \left( -2 dx^+ dx^- - f(M_6) r^{2 z-2} (dx^{+})^2 - r^{z-2}C dx^+ + d{\mathbf{x}}^2 \right) + \frac{dr^2}{r^2}.
\eea
where $z$ is the dynamical exponent of the NR CFT. This metric captures all the symmetries of the Schr\"{o}dinger symmetry group for $z \neq 1$, with $z=1$, $f(M_6) = C = 0$ being the standard metric on $AdS_5$. Here $f(M_6)$ is a function of the $M_6$ coordinates - from the external space perspective it is simply a scalar. We also demand that $C$ is a one-form invariant under the isometries of $M_6$, by requiring that $\mathcal{L}_{\zeta} C = 0$, where $\zeta$ is an $M_6$ Killing vector. This metric combines some of the features considered in \cite{Hartnoll:2008rs,Donos:2009en} for providing NR CFT duals from known $AdS$ solutions. 

Given a metric $g_{\mu \nu}$, the Killing vectors $\zeta$ may be determined from the solutions of
\be
\mathcal{L}_{\zeta}g_{\mu \nu} = \zeta^{\rho} \partial_{\rho} g_{\mu \nu} + g_{\rho \nu} \partial_{\mu} \zeta^{\rho} + g_{\mu \rho} \partial_{\nu} \zeta^{\rho}.
\ee
For the above metric the Killing vectors are
\bea
&&\partial_{i} \; , \;\;\; \;\;\; \partial_{+}\; , \;\;\;\;\;\; \partial_{-}\; , \;\;\;\;\;\; x^2\partial_{1}-x^1\partial_{2}\; , \;\;\;\;\;\; x^i\partial_{-} - x^+ \partial_{i},\cr\cr
&& zx^+\partial_{+} + x^i\partial_{i} + (2-z)x^-\partial_{-} + r\partial_{r}.
\eea
When the symmetry group is enlarged to Schr\"{o}dinger symmetry, there is an extra Killing vector
\be
-(x^+)^2\partial_{+} -\frac{1}{2}\left(\frac{1}{r^2} + (x^i)^2\right) \partial_{-} - x^+ x^i\partial_{i} + x^+ r\partial_{r}.
\ee
In total these generators lead to isometries with the following infinitesimal form
\bea
 P^i &:& \;\;\delta x^i = a^i,  \quad \;H: \;\; \delta x^+ = a, \quad M :\; \delta x^- = a,  \quad M^{12} :\;\delta x^i = a \epsilon^i_{\;j} \;x^j,  \crcr  K^i &:& \; \; \delta x^i = - a^i x^+, \quad \delta x^- = a^i x^i,\\
 D&:& \;\; \delta x^i = a x^i, \quad  \delta r = a r, \quad\delta x^{+} =  za x^+ , \quad \delta x^{-} =  (2-z)a x^-, \crcr
C &:&  \;\; \delta x^{i} = - a x^+ x^i,  \quad \d r = + a x^{+} r, \quad \delta x^{+} = -a (x^{+})^2,  \quad \d x^- = - \frac{a}{2}(\frac{1}{r^2} +(x^i)^2). \nonumber
\eea
The non-trivial commutators of these generators are given by
\be
\begin{array}{ll}
 \left[M^{12},P^i\right] = i(\d^{1i}P^2-\d^{2i}P^1 ), & \left[M^{12},K^i\right] = i(\d^{1i}K^2-\d^{2i}K^1 ), \\ \left[D,K^i\right]=i(1-z) K^i,  &
 \left[D,P^i\right]=-iP^i,  \\ \left[D,H\right]=-ziH, & \left[P^i,K^j\right]=-i\d^{ij}M,
\end{array}
\ee
with the additional commutators for $z=2$
\bea
&& \left[D,M\right]=i(2-z) M, \;\;\; \left[D,C\right] = 2i C,   \;,\;\;\; \left[H,C\right]=iD. 
\eea
In terms of the generators of $SO(4,2)$ conformal group $\tilde{P}^{\mu}, \tilde{K}^{\mu}, \tilde{M}^{\mu\nu}$ and $\tilde{D}$ in light-cone coordinates, these may be expressed as
\bea
&& P^i = \tilde{P}^i,  \;\;\;\;\;\; M^{12} = \tilde{M}^{12} , \;\;\;\;\;\;M = \tilde{P}^+,    \;\;\;\;\;\; H = \tilde{P}^-, \\
&& K^i = \tilde{M}^{i+} , \;\;\;\;\;\; D = \tilde{D} + \tilde{M}^{+-} ,  \;\;\;\;\;\; C=\frac{\tilde{K^+}}{2}.
\eea

Having presented the form of the metric for general $z$, we now wish to consider the candidate forms for the four-form field strength. Given the original magnetic field strength $G_0$ \cite{Gauntlett:2004zh}, we can imagine adding electric flux $F \equiv F_{+abc}dx^{+abc}$ \footnote{$dx^{a_1..a_n} \equiv dx^{a_1} \wedge dx^{a_2} \wedge \cdots \wedge dx^{a_n}$}, so that the total flux is 
\be
G = G_0 + F. 
\ee
As there are many potential candidates for $F$, one approach to whittle down the options is to ask that $F = \sum_{n=1}^{n=4} A^{(n)} \wedge B^{(4-n)}$, where $A$ and $B$ are forms on the external and internal space respectively, which are invariant under the respective Killing vectors of these spaces. As the Killing vectors on the internal and external space commute with each other, one may consider the invariant forms separately. As the explicit solutions on $M_6$ depend on the nature of the four-dimensional base space $\mathcal{B}_4$, we postpone treatment of them until later. 

For the external metric, once the Killing vectors $\zeta$ are determined, we consider the various $n$-forms $A^{(n)}$ satisfying
\be
\mathcal{L}_{\zeta}A^{(n)}_{a_1...a_n} = 0. 
\ee
The metric (\ref{metric1}) preserves the following forms
\bea
c_1r^z dx^{+} + c_2r^{-1} dr, \quad
 r^{z-1} dx^{+r},\quad
 r^{z+2} dx^{+12}, \quad
c_3 r^{z+1} dx^{+r12} + c_4 r^{4} dx^{+-12},
\label{extform2}
\eea
where $c_{i}$ denote arbitrary constants. Note due to the different exponents of $r$ present, it is not possible for any of these terms to mix with each other under the symmetry group. For $z=2$, the surviving forms after special conformal symmetry is introduced are
\be
\label{extform}
r^2 dx^{+}, \quad r dx^{+r}, \quad r^4 dx^{+12}, \quad r^3 dx^{+r12}.
\ee
Having sketched the general scenario, we devote the rest of the paper to the analysis of the $z=2$ duals. The more general case may be tackled later. In this note, we focus on solutions to Einstein's equation and the flux equations in $d=11$. Given our ansatz, only when $C=0$ and the fluxes are purely magnetic $G=G_0$ can we preserve two supersymmetries. Following the same spinor decompositon as \cite{Gauntlett:2004zh}, $\epsilon = \psi \otimes e^{\lambda/2} \xi$, it can be shown that imposing $\rho^{+} \psi = 0$ kills all superconformal supersymmetry while preserving just two Poincar\'{e} supersymmetries. 

Coupled with studies of NR CFT duals there has been some work done on consistent truncations \cite{Maldacena:2008wh, Gauntlett:2009zw}. In passing, we mention that a consistent reduction to $d=5$ minimal gauged supergravity of these solutions which appeared in  \cite{Gauntlett:2006ai}, does not allow a deformation preserving Schr\"{o}dinger symmetry. A consistent reduction from the $AdS_5 \times_{w} M_6$ solutions in $d=11$ to $d=5$ gauged supergravity allowing massive vector fields is expected to exist. 

\section{$\mathcal{B}_{4} = S^2 \times S^2$}
As mentioned in the introduction, the $AdS_5 \times_{w} M_6$ solutions where $M_6$ is complex, have either $KE_4$ bases or product bases comprised of Riemann surfaces. When one of the Riemann surfaces are $T^2$, one can reduce to type IIA/IIB, whereas the other explicit supersymmetric solutions in the family are purely M-theoretic in nature. In this section we focus on the case $\mathcal{B}_4 = S^2 \times S^2$. In general the volumes of the two spheres are different, but when the volumes are the same, we get the $KE_4$ base $S^2 \times S^2$. As the structure of the explicit solutions is more or less the same, we feel that it is sufficient to focus on this particular case. The case $S^2 \times T^2$ which admits a $d=10$ description we consider in the next section. 

In general for $\mathcal{C}_1 \times \mathcal{C}_2$, the six-dimensional metric takes the form
\bea
ds^2(M_6) = \sum_{i=1}^{2}\frac{1}{3} e^{-6 \lambda}(a_i-k_iy^2)ds^2(\mathcal{C}_i) + e^{-6 \lambda} \sec^2 \zeta dy^2 + \frac{1}{9} \cos^2 \zeta D \psi^2,
\eea
where $ds^2(\mathcal{C}_i)$ denote the metric on $S^2$ or $H^2$, with the curvatures $k_i$ taking the appropriate values $k=+1$ or $k=-1$ accordingly. Here $D \psi \equiv d \psi + P$ with $0 \leq \psi \leq 2 \pi$ and  $d P = vol_1 + vol_2$. The original magnetic four-form flux is 
\be
\label{o4form}
G_0 = g_1 {vol}_1 {vol}_2 + g_2 {vol}_1dy D \psi + g_3{vol}_2 dy D \psi,
\ee
where ${vol}_{i}$ denotes the volume form on the product spaces. The explicit forms of $e^{6 \lambda}, \cos^2 \zeta, g_i$, which are all functions of y depending on constants $a_i, k_i$ and $c$, may be found in \cite{Gauntlett:2004zh}.

From here on, we restrict ourselves to the $c=0$ case of $S^2 \times S^2$ ($k_1 = k_2 = 1$). When $c=0$, $y$ is bounded above and below by the zeroes of $\cos^2 \zeta$
\be
\label{yrange}
y^2 \leq  \frac{1}{2 \sqrt{3}} \sqrt{3 a_1^2 + 3 a_2^2 + 10 a_1 a_2} -\frac{a_1+a_2}{2}.  
\ee
In addition to the original magentic flux (\ref{o4form}), we consider the flux 
\bea
F &=& f_1(y) r^3 dx^{r+12} - \frac{f_1'(y)}{4} r^4 dx^{+12y} + f_2(y) r dx^{r+y} D \psi + f_3(y) r dx^{r+} vol_1 \\ &+& f_4(y) r dx^{r+} vol_2
+ \frac{1}{2}[f_2(y)-f_3(y)']r^2 dx^{+y} vol_1 + \frac{1}{2} [f_2(y)-f_4(y)'] r^2 dx^{+y} vol_2, \nonumber
\eea
which is constructed from forms invariant under the desired internal and external symmetries. It is also by construction closed $d F = 0$, so that the Bianchi is satisfied. 

In tandum with these fluxes, we consider the invariant one-form $C$ to be of the form 
\be
C \equiv C_{y}(y) dy + C_{\psi}(y) D \psi,  
\ee
thus exhausting the permitted invariant one-forms on the original $M_6$.  

We now turn to the flux equations of motion. Ensuring that the equation of motion is satisfied leads to five equations where both $r$ and $C_y$ drop out: (\ref{prodflux1}), (\ref{prodflux2}), (\ref{prodflux3}), (\ref{prodflux4}) and (\ref{prodflux5}). To avoid clutter we reproduce these and subsequent equations of motion in the appendix. 

We stress again that any solution of $d=11$ supergravity necessitates that these be solved along with Einstein equation if a solution is to exist. Note, not all of the above are independent: taking the derivative of (\ref{prodflux1}), while making use of (\ref{prodflux2}) and (\ref{prodflux3}), plus $g_{1}'=g_2 +g_3$, (from Bianchi, see (\ref{o4form})) we find (\ref{prodflux4}).

We now proceed step by step. We take $f(M_6)$ to be just a function of $y$ for simplicity $f(M_6) \equiv f(y)$. This has the upshot that none of the internal isometries are spoiled.
From the equations of motion, it is clear that $C_{\psi}$ can only appear when flux $f_i$ terms are switched on. 
 
Bearing in mind that the original background \cite{Gauntlett:2004zh} satisfies Einstein equations $E_{AB} = 0$, the introduction of $f(y), C_{\psi}$ and fluxes $f_i$ make some components non-zero. When $C_{\psi}$, and $f_i$ are zero, $g_{++} \sim f(y)$ and one may simply determine $f(y)$ by solving $E_{++} =0$. The result for $a_1=a_2=1$ is 
\bea
f &=& \alpha_1 y + \alpha_2 \biggl(-3 y \tan^{-1} \left[\frac{1}{3} \sqrt{9+6 \sqrt{3}}(-3+2 \sqrt{3})y \right] \nn &+& (45y \sqrt{3} + 78 y)\tanh^{-1}\left[\frac{1}{3} \sqrt{9+6 \sqrt{3}} \sqrt{3} y \right] - 4 \sqrt{9+6 \sqrt{3}}(2+\sqrt{3}) \biggr),
\eea
where $\alpha_i$ are integration constants. What is important to note, is that $f$ changes sign as y approaches the bound (\ref{yrange}). This means that in principal it is susceptible to instabilities which we address in the appendix.

However, when one attempts to have non-zero $f_i$ but $C_{\psi} =0$, it is satisfying to see that no solution exists. In addition to the four-independent flux equations, Einstein imposes
\be
f_2 = \frac{1}{2} ( f_3' + f_4'), 
\ee
and either $f_3=f_4$ \textit{or} $a_1 = a_2$ is imposed. The combination of (\ref{prodflux2}) and (\ref{prodflux3}) then demand that both $f_3 = f_4$ and $a_1=a_2$ with the remaining equations only being satisfied if all $f_i = 0$. 

In general when $C_{\psi} \neq 0$, after considerable numerical work for $a_1 \neq a_2$, one can show by expanding in terms of power series that given our ansatz, the only  supergravity preserving Schr\"odinger symmetry and the internal symmetry of $M_6$ has trivial electric field strengths $f_i = 0$.

Finally, we remark that when $a_1=a_2$, the base becomes $KE_4$, so we do not expect there to be any solutions with $KE_4$ base when $C_{\psi}$ and $f_i$ are non-zero. 

\section{$\mathcal{B}_4 = S^2 \times T^2$ : Sasaki-Einstein}
We next consider the family of solutions in \cite{Gauntlett:2004zh} with $\mathcal{B}_4 = S^2 \times T^2$. By dimensional reduction and T-duality, these solutions are related to type IIB solutions with Sasaki-Einstein five-manifolds $AdS_5 \times S_5$ \cite{Gauntlett:2004yd}. The NRCFT duals of these solutions have already been discussed in \cite{Herzog:2008wg,Maldacena:2008wh,Adams:2008wt,Hartnoll:2008rs}, so in this section we hope to provide the $d=11$ angle on the story. 

The metric is given by
\bea
ds^2(M_6) = e^{-6 \lambda} ds^2(T^2) + \frac{1-cy}{6} ds^2(S^2) + e^{-6 \lambda}\sec^2 \zeta dy^2  + \frac{1}{9} \cos^2 \zeta D \psi^2, 
\eea
where $(x_3,x_4)$ parametrise our torus and $(\theta,\phi)$ a unit radius two-sphere. 

As before $\psi$ has period $2 \pi$ and this time $D \psi \equiv d \psi + P$ with $d P = vol(S^2)$. Here $a \in (0,1)$ and without loss of generality, we may take $c=1$. The roots of $\cos^2 \zeta$ define a range for y: $y_1 \leq y \leq y_2$. The parameter $a$ may be fixed in terms of two relatively prime integers $p> q > 0$ \cite{Gauntlett:2004yd},
\be
a = \frac{1}{2} + \frac{3 q^2-p^2}{4 p^3} \sqrt{4 p^2 - 3 q^2}. 
\ee
In terms of these integers the roots are
\bea
\label{ypqroot}
y_1 &=& \frac{1}{4p}(2p -3q - \sqrt{4 p^2 - 3 q^2}), \nn
y_2 &=& \frac{1}{4p}(2p +3q - \sqrt{4 p^2 - 3 q^2}). 
\eea
The original four-form flux is 
\be
G = g_1 dx^{34} vol(S^2) + g_2 vol(S^2) d y D \psi + g_3 dx^{34y} D \psi,
\ee
where the explict forms of $e^{6 \lambda}, \cos^2 \zeta, g_i$ are recoverable in \cite{Gauntlett:2004zh}.

Proceeding as in the previous section, the most general electric flux satisfying the Bianchi and being invariant under the isometries is
\bea
F &=& f_1(y) r^3 dx^{r+12} - \frac{f_1(y)'}{4} r^4 dx^{+12y} + f_2(y) r dx^{r+34} + f_3(y) r dx^{r+} Vol(S^2) \nn &+& f_4(y) r dx^{r+y} D \psi
+ f_5(y) r dx^{r+3} D \psi + f_6(y) r dx^{r+4 } D \psi - \frac{f_2(y)'}{2} r^2 dx^{+34y} \nn &+& \frac{1}{2}[f_4(y) - f_3(y)'] r^2 dx^{+y} Vol(S^2) + \frac{f_5(y)}{2} r^2 dx^{+3} Vol(S^2) + \frac{f_5(y)'}{2} r^2 dx^{+3y} D \psi \nn &+& \frac{f_6(y)}{2} r^2 dx^{+4} Vol(S^2) + \frac{f_6(y)'}{2} r^2 dx^{+4y} D \psi,
\eea
with the resulting equations of motion again appearing in the appendix. 

Again using the Bianchi $g_1'=g_3$, one can show that the derivative of (\ref{SEflux1}) with (\ref{SEflux3}) means that (\ref{SEflux4}) is trivially satisfied. Also $f_5$ and $f_6$ decouple from the rest of the other $f_i$ and from each other, though they are the same up to a change in sign. We proceed parallel to the last section. 

In the absence of one-form $C \equiv C_{\psi} D \psi + C_{3} dx^{3} + C_4 dx^{4}$, and the $f_i$, one can solve for $f(M_6) \equiv f(y)$. The solution is 
\bea
f(y) &=& \alpha_1 y + \alpha_2 y \left[-\frac{1}{ay} -\frac{1}{a} \sum_{roots} \frac{(3-2 yi) \ln(y-y_i)}{6 y_i (1-y_i)} \right],
\eea
with $\alpha_i$ being integration constants. Here the latter part is not real over the range of $y$, so $\alpha_2$ is necessarily zero. This leaves $f(y) \sim y$, and as $y$ changes sign, so does $f(y)$. The stability of the dimensionally reduced IIB solution was treated in $\cite{Hartnoll:2008rs}$, where it was shown that the solution is unstable provided the fluxes $f_i$ are sufficiently small $f_i$. The same conclusion may be applied here also.  

The first indication that something special has happened when the $S^2$ of the last section is replaced by $T^2$, is the decoupling of $f_5$ and $f_6$. These should correspond to the type IIB solutions presented in \cite{Herzog:2008wg,Maldacena:2008wh,Adams:2008wt,Hartnoll:2008rs}, which were largely the result of TsT transformations. Here TsT refers to the process of T-dualising on an internal $U(1)$, say $\psi$, shifting along $x^{-}$, $x^{-} \rightarrow x^{-} + \sigma \psi$, then T-dualising back along $\psi$. 

We uplift these solutions in appendix C and they act as a consistency check for our flux equations of motion. In terms of our $f_i$, these solutions take the form of one of the following,
\bea
\label{soln1}
f_5 &=& -\frac{2 \sigma_1 \cos^2 \zeta}{3}, \quad C_4 = 2 \sigma_1 y, 
\mbox{ \textbf{or} }
f_6 = -\frac{2 \sigma_2 \cos^2 \zeta}{3}, \quad C_3 = -2 \sigma_2 y,  
\eea
depending on which cycle of the torus $T^2$ one uplifts on. One can then complete the solution by solving the $E_{++}=0$ component for $f(y)$. In contrast to IIB, in the M-theory setting, we can have both terms simultaneously as they decouple from each other. 

Bearing in mind that the Killing vector $\partial_{\psi}$ of the $AdS_5 \times_{w} M_6$ solutions is a linear combination of the type IIB Reeb vector $\partial_{\psi'}$ with a torus one-cycle, say $x_3$, $\partial_{\psi} = \partial_{\psi'} + \partial_3$, when one reduces to IIA and performs a TsT on $\partial_{\psi}$, one recovers the same flux above, but with a different value of $C_{3}$. In fact for $f_{6} = \sigma \cos^2 \zeta$, one can integrate (\ref{SEflux6}). The solution is 
\be
C_{3} = 3 \sigma \frac{(a-y)}{(1-y)} + c \frac{(a-y^2)}{(1-y)},
\ee
for arbitrary constant $c$. This is also a solution to Einstein's equation. This highlights the freedom presented when one performs the TsT transformation, as one can choose a linear combination of the two $U(1)$'s from type IIB or IIA. For example, (\ref{soln1}) corresponds to the case $c=2$ when $\sigma$ is properly rescaled.   

Also possible, is the reduction to type IIA and the TsT using $x^{-}$ and $x^{3}$. In this case the solution becomes 
\bea
f_1 &=& -4 \sigma, \quad f_2 = 2 \sigma e^{-6 \lambda}, \quad C_{\psi} = - \frac{3 \sigma }{2} \frac{-2 y + y^2 +a}{a-3y^2+2 y^3},.
\eea

Away from the solutions which may be generated in such a fashion, one may tackle Einstein's equation and flux equations head on by assuming that 
all the $f_i$'s are analytic in the interval (\ref{ypqroot}) and may be expanded in a power series in $y$. This admits solutions seeded by four integral constants. From the form of the fluxes, these constants are related to the charges of M2 branes stretched along different two-cycles in the geometry. The general solution can be presented as a power series, where the coefficients are determind by those four integral constants. Unfortunately we were unable to find a neat expression for these solution, so we merely document their existence.  

\section{Discussion}
The subject of this work was M-theory solutions admitting $d=2+1$ non-relativistic conformal field theory duals. This necessitates a five-dimensional holographic dual. Our starting point was the explicit $AdS_5 \times_{w} M_6$ solutions presented in \cite{Gauntlett:2004zh}, before  replacing the external $AdS_5$ space with a space with the reduced symmetries of the Schr\"{o}dinger algebra. 

In section 2, en route to constructing the candidate fluxes, we considered the most general forms preserving the required symmetries. In section 3, we analysed the possibility of solutions where $M_6$ has base space $\mathcal{B}_{4} = S^2 \times S^2$. We found that the only supersymmetric solution consists of adding $g_{++}$ scalar function $f(y)$ to the metric in the presence of the original magnetic fluxes. As this family has overlap with $\mathcal{B}_{4} = KE_4$, we do not expect any non-trivial flux solutions here either. Owing the the similar structure of the $AdS_5 \times_{w} M_6$, we also expect this result to hold for all products without a $T^2$ metric in the base. 

In section 3, we replace one of the $S^2$ with $T^2$ and reconsider the problem. The two one-cycles of $T^2$ allow us to consider a more general flux ansatz, where $x^3$ and $x^4$ appear separately in the candidate electric four-forms. In fact these extra flux terms decouple from the other fluxes, and we are able to find a family of solutions. These appear in \cite{Herzog:2008wg,Maldacena:2008wh,Adams:2008wt,Hartnoll:2008rs}, but here we can now add a second shifting parameter as we approach the problem directly from $d=11$ supergravity. These solutions also have a second clear $U(1)$ appearing as a candidate for TsT transformation, allowing yet another generalisation of solutions in \cite{Herzog:2008wg,Maldacena:2008wh,Adams:2008wt,Hartnoll:2008rs}. 

In future, we would like to return to these backgrounds but focus more on supersymmetry and extend the search for solutions at different $z$, thus generating the results on the Schr\"{o}dinger symmetry appearing here. One way to pursue this, would be investigating the possibility of consistent truncations on $M_6$ from $d=11$ to a $d=5$ gauged supergravity theory with a massive vector field. 
This would extend some of the work of \cite{Maldacena:2008wh, Gauntlett:2009zw}, which was performed for $SE_5$ and $SE_7$, where also lower bounds were given for supersymmetric solutions.  

\section*{{\large Acknowledgements}}

We are grateful to Jerome Gauntlett, Ki-Myeong Lee, Sangmin Lee, Sungjay Lee, Oscar Varela and Patta Yogendran for discussion. 
\appendix

\section{Equations of motion}
Here we house the constraints on the candidate flux terms $f_i$ arising from the equations of motion. 

\noindent{$\mathbf{S^2 \times S^2}$ \\
In this case the constraints are
\bea
\label{prodflux1}
f_1 g_1 &=&  \frac{4}{3e^{6 \lambda}} f_2 (a_1 - k_1 y^2)(a_2 - k_2 y^2) -\frac{1}{6}e^{6 \lambda} \cos^2 \zeta \frac{(a_2-k_2 y^2)}{(a_1-k_1 y^2)}[f_2 - f_3'+ 2 C_{\psi} g_2] \nn &-& \frac{1}{6}e^{6 \lambda} \cos^2 \zeta \frac{(a_1-k_1 y^2)}{(a_2-k_2 y^2)}[f_2-f_4' + 2 C_{\psi} g_3], \\
\label{prodflux2}
f_1 g_2 &=&  \frac{4}{3} f_4 \frac{(a_1 - k_1 y^2)}{(a_2 - k_2 y^2)} - \frac{1}{6}\left[e^{6 \lambda} \cos^2 \zeta \frac{(a_1-k_1 y^2)}{(a_2-k_2 y^2)}[f_2-f_4'+2 C_{\psi} g_3]\right]', 
\eea
\bea
\label{prodflux3}
f_1 g_3 &=& \frac{4}{3} f_3 \frac{(a_2 - k_2 y^2)}{(a_1 - k_1 y^2)} - \frac{1}{6}\left[e^{6 \lambda} \cos^2 \zeta \frac{(a_2-k_2 y^2)}{(a_1-k_1 y^2)}[f_2-f_3'+ 2 C_{\psi} g_2]\right]', \\
\label{prodflux4}
\frac{1}{4}f_1' g_1 &=&   \left[\frac{f_2}{3 e^{6 \lambda}} (a_1 - k_1 y^2)(a_2 - k_2 y^2)\right]' - \frac{1}{3} f_3 \frac{(a_2 - k_2 y^2)}{(a_1 - k_1 y^2)} - \frac{1}{3} f_4 \frac{(a_1 - k_1 y^2)}{(a_2 - k_2 y^2)}, \\
\label{prodflux5}
0 &=& f_2 g_1 + f_3 g_3 + f_4 g_2 - \frac{2}{27} \frac{f_1}{e^{12 \lambda}} (a_1 - k_1 y^2)(a_2 - k_2 y^2) \nn &-& \frac{1}{4(27)}\left[\frac{f_1' \cos^2 \zeta}{e^{6 \lambda}} (a_1 - k_1 y^2)(a_2-k_2 y^2) \right]',
\eea
where dashes denote derivatives with respect to $y$. 

\noindent 
$\mathbf{S^2 \times T^2}$ \\
Here the equations of motion are 
\bea
\label{SEflux1}
0 &=& f_1 g_1 - 2 f_4 (1-y) + [f_4 - f_3'+2 g_2 C_{\psi}] \left(\frac{\cos^2 \zeta}{1- y}\right) , \\
0 &=& f_1 g_2 - \frac{2}{9} f_2 e^{6 \lambda} (1- y) - \frac{1}{36} \left[(f_2' -2 g_3 C_{\psi})\cos^2 \zeta e^{12 \lambda} (1- y) \right]', \\
\label{SEflux3}
0 &=& f_1 g_3 - \frac{8 f_3}{(1-y)e^{6 \lambda}} + \left[[f_4-f_3'+2 g_2 C_{\psi})] \left( \frac{\cos^2 \zeta}{1-y}\right) \right]', \\
\label{SEflux4}
0 &=& -\frac{f_1' g_1}{4} - \frac{2 f_3}{(1-y) e^{6 \lambda}} + \frac{1}{2} [ f_4 (1-y)]', \\
0 &=& -2 f_5 \left( \frac{1-y}{\cos^2 \zeta} \right)- \frac{1}{4}[(f_5'+2 C_4 g_3) e^{6 \lambda} (1- y)]' + \frac{f_5+2 C_4 g_1}{1-y}, \\
\label{SEflux6}
0 &=& 2 f_6 \left( \frac{1- y}{\cos^2 \zeta} \right) + \frac{1}{4}[(f_6'-2 C_3 g_3) e^{6 \lambda} (1- y)]'- \frac{f_6-2 C_3 g_1}{1-y}, \\
0 &=& f_2 g_2 + f_3 g_3 + f_4 g_1 - \frac{f_1}{9} \frac{1-y}{ e^{6 \lambda}}- \frac{1}{72}[f_1' \cos^2 \zeta (1-  y)]'.
\eea

\section{Stability analysis}
We begin by assuming that the original $AdS_5 \times_{w} M_6$ is stable and consider for simplicity the variation in the external metric $\delta g_{\mu \nu} \equiv h_{\mu \nu}$, where $h$ is traceless $g^{\mu \nu} h_{\mu \nu} = 0$ and transverse $d \star h = 0$. We ignore the tensor modes on the internal manifold as $M_6$ is unaffected when deforming to a Sch($2$) geometry. Following discussions in \cite{DeWolfe:2001nz}, the first order change in Ricci tensor under this perturbation is then given by the Lichnerowicz operator
\bea
\delta R_{ab} &=& \tfrac{1}{2} (\Delta_{L} h)_{ab}, \nn
(\Delta_{L} h)_{ab} &=& 2 R^{c}_{~abd}h^{d}_{~c} + R_{ca} h^{c}_{~b} + R_{cb} h^{c}_{~a} - \nabla^{c} \nabla_{c} h_{ab}, 
\eea
with the requirement that \be \delta R_{ab} = -4 h_{ab}. \label{constraint} \ee This property follows from the Ricci tensor of the original $AdS_5$ solution $R_{ab} = - 4 g_{ab}$. Combining these, we get the equation 
\be
- h_{ab} - \nabla^{c} \nabla_{c} h_{ab} = 0. 
\ee

Next we can expand the tensor mode in a parallel fashion to \cite{Hartnoll:2008rs}
\be
h_{ab} = \tilde{h}_{ab}(r) Y(M_6) e^{-i \omega x^{+} + i k \cdot x -i M x^{-}},
\ee
leading to the equations 
\bea
[-\nabla^2_{M_6} + M^2 f] Y - \lambda Y &=& 0, \nn
\left[ \frac{d^{2} }{d^{2} r} + \frac{5}{r} \frac{d}{dr} + \frac{2 M \omega - k^2}{r^4} - \frac{\lambda-1}{r^2} \right] \tilde{h}_{ab}(r) &=& 0,
\eea
where the negative contribution to the mass term comes from the lower bound on Lichnerowicz spectrum $-2 d$ for $AdS_d$. At this point, we can then borrow the analysis presented in \cite{Hartnoll:2008rs} where it is concluded that large $M$ will lead to negative $\lambda$, meaning that the energy of the above system will not be bounded below. We conclude that this solution is unstable. 

\section{Type IIB solution uplift }
Here we begin with the IIB solution in \cite{Hartnoll:2008rs} with $B$-field and uplift to M-theory. 

Employing a shift, $\beta = -6 x_3 - c \psi$ \cite{Gauntlett:2004yd}, the metric may be re-written as in \cite{Gauntlett:2004zh}
\bea
ds^{2} &=& \biggl[ r^2 ( -2 dx^+ dx^- - f(X_5) r^2 (dx^{+})^2 + d\mathbf{x}^2) + \frac{dr^2}{r^2} + \frac{1-cy}{6} (d \theta^2 + \sin^2 \theta d \phi^2) \nn &+&
e^{-6 \lambda} \sec^2 \zeta dy^2 + \frac{\cos^2 \zeta}{9}D \psi^2 \nn &+& e^{6 \lambda} [d x_3 + \frac{-2y + y^2 c + a c}{6(a-y^2)}D \psi]^2
\biggr], \nn
B_{2} &=& \sigma r^2 dx^{+} \wedge \frac{1}{3}[(1-cy)D \psi - 6ydx_3], \nn
F_{5} &=& 4(1+\star)\mbox{vol}(X_5).
\eea
After T-duality \cite{Hassan:1999mm} on $x_3$, and subsequent uplifting to M-theory on $x_4$, the final solution is 
\bea
ds^2 &=& e^{2 \lambda} \biggl[ r^2 (-2 dx^{+} dx^{-} - f(X_5)r^2 (dx^{+})^2 +  d\mathbf{x}^2) + \frac{dr^2}{r^2} \nn &+& 
e^{-6 \lambda} (dx_4^2 + [dx_3 - 2y \sigma r^2 dx^{+}]^2) + \frac{1-cy}{6}ds^2(S^2)+ e^{-6 \lambda} \sec^2 \zeta dy^2 + \frac{1}{9} \cos^2 D \psi^2 \biggr], \nn
A^{(3)} &=& \frac{-2 y + y^2 c + ca}{6(a-y^2)} dx^{3} dx^{4} D \psi - \frac{\sigma r^2  \cos^2 \zeta}{3} dx^{+} dx^{4} D \psi.  
\eea

\end{document}